\documentclass[iop]{emulateapj}
\usepackage{amsmath}

\newcommand{\Msolar}{M$_{\odot}$}
\newcommand{\Lsolar}{L$_{\odot}$}

\begin{document}

\title{Consequences of Dynamical Disruption and Mass Segregation for the Binary Frequencies of Star Clusters}
\shorttitle{Dynamical Disruption and Mass Segregation of Binaries in Star Clusters}

\author{Aaron M.~Geller\altaffilmark{1,2},
Richard de Grijs\altaffilmark{3,4},
Chengyuan Li\altaffilmark{4,3}
and Jarrod R.~Hurley\altaffilmark{5} \\
\scriptsize{
\textup{
\affil{1}{$^1$Center for Interdisciplinary Exploration and Research in Astrophysics (CIERA) and Department of Physics and Astronomy, Northwestern University, 2145 Sheridan Rd, Evanston, IL 60208, USA; a-geller@northwestern.edu} \\
\affil{2}{$^2$NSF Astronomy and Astrophysics Postdoctoral Fellow} \\
\affil{3}{$^3$Kavli Institute for Astronomy and Astrophysics, Peking University, Yi He Yuan Lu 5, Hai Dian District, Beijing 100871, China} \\
\affil{4}{$^4$Department of Astronomy, Peking University, Yi He Yuan Lu 5, Hai Dian District, Beijing 100871, China} \\
\affil{5}{$^5$Centre for Astrophysics and Supercomputing, Swinburne University of Technology, VIC 3122, Australia}}}}

\shortauthors{Geller et al.}

\begin{abstract}

The massive (13,000--26,000 \Msolar), young (15--30 Myr) Large
Magellanic Cloud star cluster NGC~1818 reveals an unexpected
increasing binary frequency with radius for F-type stars (1.3--2.2
\Msolar). This is in contrast to many older star clusters that show
a decreasing binary frequency with radius. 
We study this phenomenon with sophisticated
$N$-body modeling, exploring a range of initial conditions, from
smooth virialized density distributions to highly substructured and
collapsing configurations.  We find that many of these models can
reproduce the cluster's observed properties, 
although with a modest preference for substructured 
initial conditions. 
Our models produce the observed radial trend in binary frequency through
disruption of soft binaries (with semi-major axes, $a \gtrsim 3000$ AU),
on approximately a crossing time ($\sim 5.4$ Myr), preferentially in
the cluster core. Mass segregation subsequently causes the binaries to
sink towards the core. After roughly one initial half-mass relaxation
time ($t_{\rm rh}(0) \sim 340$ Myr) the radial binary frequency
distribution becomes bimodal, the innermost binaries having already
segregated towards the core, leaving a minimum in the radial binary
frequency distribution that marches outwards with time. After 4--6
$t_{\rm rh}(0)$, the rising distribution in the halo disappears,
leaving a radial distribution that rises only towards the core. Thus,
both a radial binary frequency distribution that falls towards the
core (as observed for NGC~1818) and one that rises towards the core
(as for older star clusters) can arise naturally from the same
evolutionary sequence owing to binary disruption and mass segregation
in rich star clusters.

\end{abstract}


\section{Introduction}

Observations of star-forming regions, young and old star clusters, and the Galactic field indicate that most stars reside in binary or 
higher-order multiple systems \citep[e.g.][]{kou05,kou07,rag10,gel10,kra11,gel12,kin12}.  
Since most stars (with masses $\gtrsim0.5$~\Msolar) form in clusters or groups \citep{lad03}, many of which quickly dissolve to 
populate the Galactic field \citep{ada01}, connecting observed binary frequencies within these different environments is critical to our 
understanding of star formation.
In dense star clusters, the frequency of binary stars can be significantly modified by close encounters with other stars, and these processes can 
be studied in detail through sophisticated $N$-body simulations.    

In most star clusters, the rates of binary disruption (particularly for wide binaries) greatly exceed the rates of binary creation.  
As explained by \citet{heg75}, binaries that have low binding energies relative to the kinetic energies of stars within the cluster, 
known as ``soft'' binaries, tend to become even less bound (``softer'') on average as a result of stellar interactions, and are eventually disrupted.  
Conversely binaries with high binding energies 
relative to the kinetic energies of stars within the cluster (``hard'' binaries)
become more tightly bound (``harder'').   Most hard binaries are unlikely to be disrupted by stellar encounters, but  
dynamical hardening and binary evolution processes can also destroy very hard binaries.  
These dynamical processes are expected to proceed most rapidly in the denser core of the cluster where stellar encounters are most frequent.  

Offsetting these disruption processes is the preferential tidal stripping of low-mass single stars from the cluster by the Galactic potential
and their ejection from the core due to dynamical encounters,
which, after an initial stage of rapid soft binary disruption, can lead to a roughly constant global binary frequency over many Gyr \citep{hur05,gel13}.

Locally, at different radii within a cluster, the binary frequency also evolves owing to two-body relaxation processes and dynamical friction.
On average, binaries have a higher total mass than their single-star counterparts, and therefore  energy exchange between these 
two groups tends to cause binaries to sink towards the core of the cluster.  This dynamical mass segregation results in the binary frequency 
rising in the cluster core and falling in the halo.
Indeed, many Milky Way open and globular clusters show a rising binary frequency towards the cluster core, which is interpreted to be the result of 
mass segregation \citep[e.g.][]{mat86,gel12,mil12}.

Binary disruption and mass segregation processes compete to determine the radial distribution of the binary frequency within a dense star cluster.  
Observations of star clusters of different ages allow us to study empirically the timescales for these processes and to verify the 
predictions from $N$-body simulations.
The rich star cluster NGC~1818, located in the Large Magellanic Cloud (LMC), is the youngest rich cluster where the radial dependence of the binary 
frequency has been measured \citep{els98,deg13,li13}, and is therefore very important for our understanding of the early evolution of binaries in star clusters.  

NGC~1818 has an age of 15 - 30 Myr, a total mass of 13,000 \Msolar\ to 26,000 \Msolar,
a central surface mass density of $180 \pm 4$ \Msolar~pc$^{-2}$ \citep{deg02,mac03}, and a total binary frequency\footnote{We define the 
binary frequency as $f_{\rm b} = N_{\rm b} / (N_{\rm s} + N_{\rm b} + ...)$, where $N_{\rm b}$ is the number of binaries, $N_{\rm s}$ is the number of single stars, and ``...'' signifies 
higher-order multiples.} ($f_{\rm b}$)  estimated to be between 
55\% and 100\% for F-type stars with masses between 1.3 \Msolar\ and 1.6 \Msolar\ \citep{hu10}.  
Importantly, the binary frequency in NGC~1818 is observed to decrease towards the core of the cluster \citep{deg13,li13}, in 
stark contrast to the more typically observed radial dependence where the binary frequency rises towards the core, as also predicted from 
mass segregation processes.

\citet{deg13} and \citet{li13} suggest that this radial trend of the binary frequency observed in NGC~1818 may result from the processes of binary disruption.
In this paper we test this hypothesis through sophisticated $N$-body modeling of the cluster.  In Section~\ref{method}, we describe the 
simulation method and define our grid of cluster simulations.  In Section~\ref{phenom}, we discuss the evolution of the radial dependence of 
the binary frequency and the contribution of dynamical binary disruption and mass segregation.  Then, in Section~\ref{simtoobs}, we compare 
the simulations directly with the observations and discuss our ability to reproduce the observed trend in binary frequency within our models.
In Section~\ref{discuss} we discuss the implications of these results for the origins of the radial binary frequency distribution in NGC~1818
and relate this result to similar observations in other rich star clusters.  Finally, in Section~\ref{conc} we provide our conclusions.

\section{Simulation Method} \label{method}

We use the \texttt{NBODY6} code \citep{aar03} to model the dynamical evolution of
rich star clusters, with the goal of reproducing the observations of NGC~1818 at the cluster's age.  \texttt{NBODY6} includes stellar and binary 
evolution \citep{hur00,hur02}, models dynamical encounters with binaries in detail, including those 
leading to the disruption of binaries, and mass segregation arises naturally in \texttt{NBODY6} from two-body relaxation processes.
Much of our method is identical to that of \citet{gel13}, except that here we choose some different initial 
conditions for the cluster, which we describe below.

Observations of NGC~1818 suggest that the cluster has a total mass between 13,000~\Msolar\ and 26,000~\Msolar~\citep{deg02,mac03}.
We choose to begin all of our models with 36,000 stars chosen from a \citet{kro01} IMF, with masses between 0.1~\Msolar\
(approximately the hydrogen-burning limit and the lowest stellar mass with detailed models guiding the stellar evolution code) and 50~\Msolar,
which produces an initial cluster mass of 
$\sim$21,800~\Msolar.  Extrapolating from the results of \citet{hur05} and \citet{gel13} suggests that this initial mass will remain
within the observed mass range by an age of 30 Myr (the maximum age estimate for the cluster), despite mass loss from stellar evolution, dynamical ejections and 
tidal stripping from the Galactic potential, and indeed our evolved NGC~1818 models confirm this result.
We do not model the embedded phase of the cluster here. Instead we begin our simulations at $t=0$ after gas expulsion 
and with all stars on the zero-age main sequence.

\citet{mac03} fit the observed surface brightness distribution of NGC~1818 with an EFF profile \citep{els87}, defined by
\begin{equation}
\mu(r) = \mu_0 \left(1 + \frac{r^2}{a^2}\right)^{-\gamma/2},
\end{equation}
with a central surface brightness of $\log \mu_0$~=~3.35~$\pm$~0.02~\Lsolar~pc$^{-2}$, $a$~=~12.50~$\pm$~0.78~arcsec (3.04~$\pm$~0.19~pc, 
using the canonical LMC distance modulus of 18.5, which 
equates to a scale of 4.116~arcsec~pc$^{-1}$, as in \citealt{mac03}) and $\gamma$~=~2.76~$\pm$~0.12.
Also, for reference, one can calculate the King core radius from the EFF parameters as,
\begin{equation}
r_{\rm c} = a(2^{2/\gamma}-1)^{1/2},
\end{equation}
which here equates to 2.45~pc.
\citet{mac03} adopt a mass-to-light ratio of 0.08, which implies a 
central mass surface density of 180~$\pm$~4~\Msolar~pc$^{-2}$.
We aim to reproduce these observations at the age of NGC~1818
within our simulations.

The EFF model is similar to a \citet{kin66} or \citet{plu11} model (which are more typical initial conditions for $N$-body star cluster simulations) except in the halo, where the EFF model 
maintains a slightly higher surface brightness and results in a slightly more extended cluster.  
For simplicity, here we begin our simulations with stars distributed according to a Plummer model, and we describe the details of these models below.  
At the age of NGC~1818, these 
simulations are consistent with the EFF model to within the radial extent of the \citet{deg13} and \citet{li13} studies (see Figure~\ref{sdens}).

Observations suggest that many clusters may form with subvirial velocities \citep{per06,and07,tob09,pro09}.
Therefore we investigate different simulations with initial virial ratios of 
$Q$ = 0.5 (equilibrium) and $Q$ = 0.3 and 0.1 (collapsing). 
In order to determine the initial length scale, we first ran simulations with a range of initial virial radii for each $Q$ value and determined the initial 
virial radius for a given $Q$ that best reproduces the observed surface density profile at the age of NGC~1818.  We find that an equilibrium model 
reproduces the observations with an initial virial radius of 7 pc (equivalent to an initial half-mass radius of $r_{\rm h}(0) = 5.38$~pc and 
a Plummer scale radius of $r_{\rm pl}(0) = 4.12$~pc ).
A $Q = 0.3$ model also reproduces the observations with an initial virial radius of 7 pc.
A $Q = 0.1$ model requires an initial virial radius of 10 pc ($r_{\rm h}(0) = 7.69$~pc, $r_{\rm pl}(0) = 5.89$~pc).  
For this study, we do not simulate clusters with supervirial initial conditions,
although it is conceivable that some supervirial initial conditions may reproduce the observed surface density profile of NGC~1818 at the cluster age.

\begin{figure}[!t]
\epsscale{0.86}
\plotone{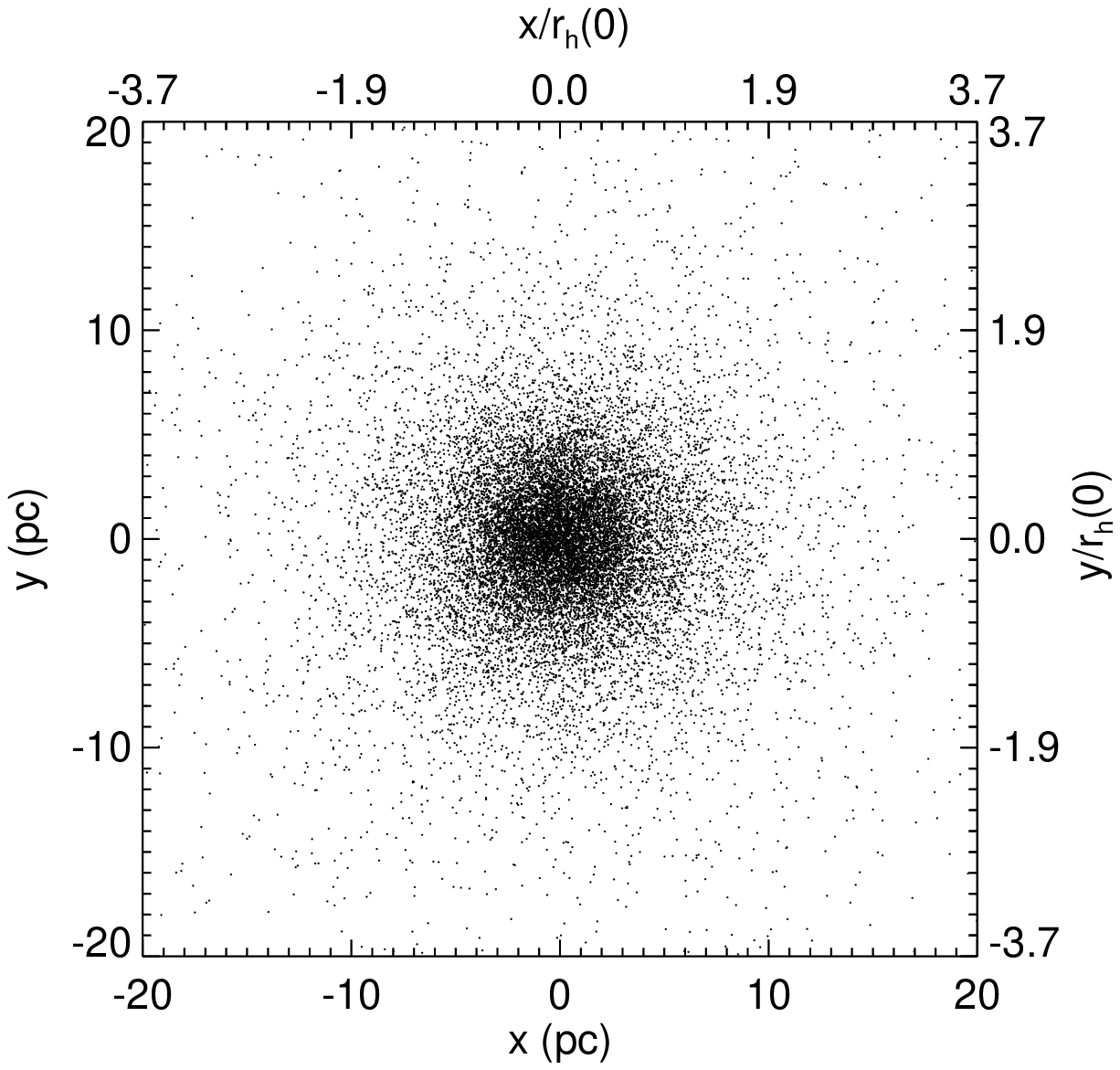}
\plotone{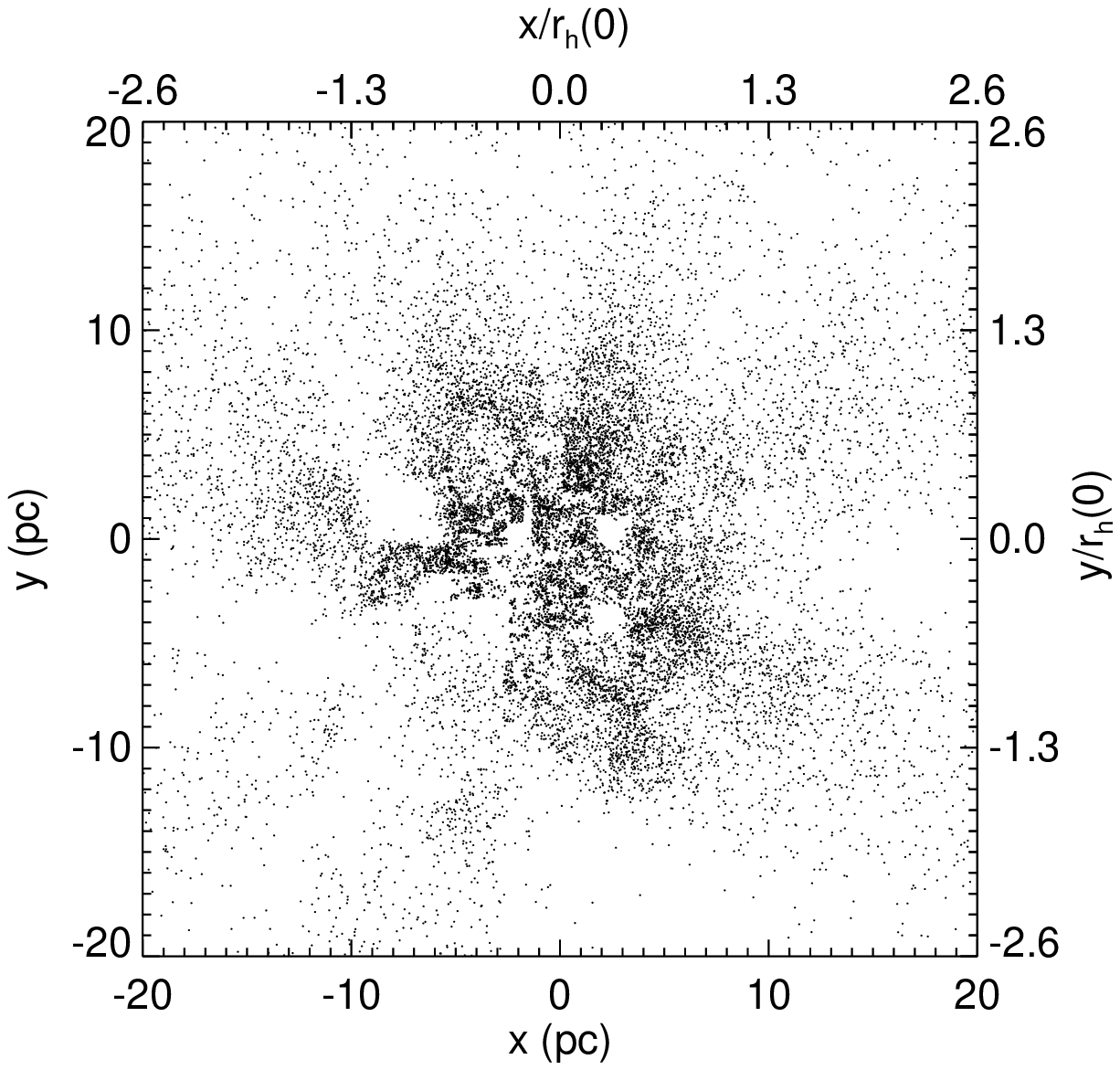}
\plotone{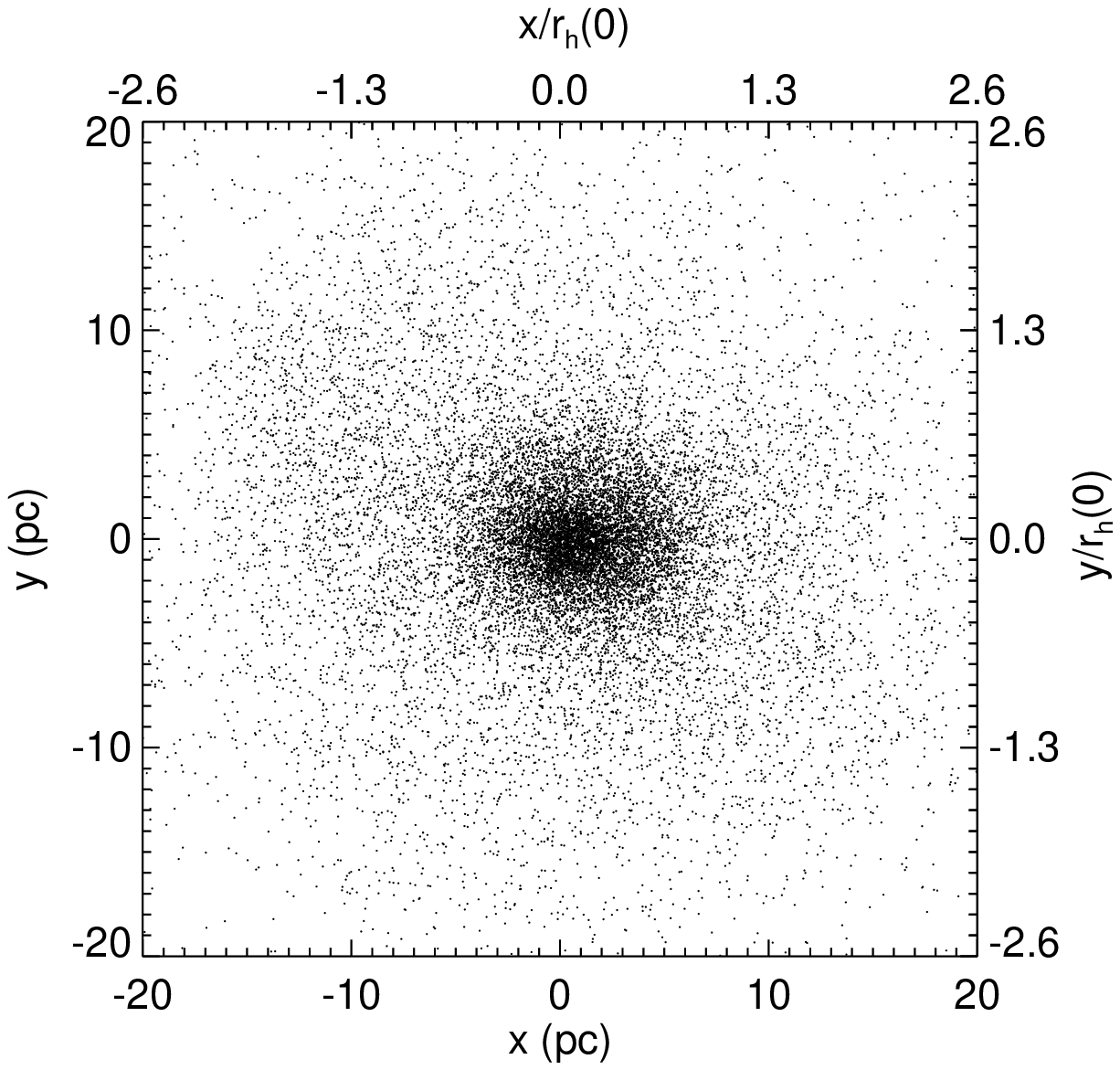}
\epsscale{1.}
\caption{
Stellar positions for one 
[$Q$,$D$] = [0.5,3.0] cluster simulation (top at $t=0$) and one 
[$Q$,$D$] = [0.3,2.0] simulation (middle at $t=0$ and bottom at $t=15$~Myr,
the minimum age estimate for NGC~1818), demonstrating the difference between smooth and clumpy (i.e. fractal) initial density distributions and 
also that the simulations with clumpy initial conditions relax to smooth density distributions by the age of NGC~1818.
The simulations are shown in projection along the $z$ axis; each simulation begins with the $x$ direction facing away 
from the LMC center, and the cluster orbits in the $x - y$ plane.  The positions are shown in pc and also in fractions of the respective initial 
half-mass radii, $r_{\rm h}(0)$.  For the 
[$Q$,$D$]~=~[0.5,3.0] simulation, $r_{\rm h}(0)$~=~5.38~pc, and for the 
[$Q$,$D$]~=~[0.3,2.0] simulation, $r_{\rm h}(0)$~=~7.69~pc.
Each binary is plotted as a single point at the center-of-mass of the given system. 
\label{xyplots}
}
\end{figure}

Many young clusters are also observed to be substructured and well represented by fractal density distributions
\citep[e.g.][]{lar95,kra08,car04,san09}.
Therefore, in addition to the smooth models at different virial ratios, we also explore different ``clumpy'' models.
We impose fractal distributions on top of the Plummer models described above (see Figure~\ref{xyplots}) 
using the McLuster code \citep{kup11}, with slight modifications to match those we made to 
\texttt{NBODY6} for defining the initial binaries \citep[see][]{gel13}, which creates initial conditions that can be easily read 
directly into \texttt{NBODY6}.  We follow \citet{goo04} and investigate clusters with fractal dimensions of $D$ = 1.6, 2.0, 2.6, and $D$ = 3.0 (which corresponds to 
the smooth models with no clumping defined above).
Again, we first ran simulations with a range of initial virial radii for each combination of $Q$ and $D$ to identify the initial length scale 
that best reproduces the observed surface density profile at the cluster age.

To define the stellar velocities for the fractal initial conditions, McLuster first draws random velocities from a normal distribution such that the 
mean motion of the stars in each subgroup is zero and the mean magnitude of the velocities for stars in each subgroup is unity.
In addition to this normally distributed velocity, the velocity of each ``parent'' is added to the velocities of each of its ``children'', so each generation 
as an ensemble moves like its parent.  Finally, the velocity of each star is multiplied by the circular velocity from the Plummer model at the star's radius.  
Thus, since the velocities are normally distributed around the parents'  (and grandparents') velocities, the initial velocities are coherent but random.  
There is evidence that most (and perhaps all) substructured clusters form with subvirial velocities \citep{gir12}, and furthermore
in order to erase the initial substructure as rapidly as is observed, subvirial initial conditions may be required \citep{goo04}.
For our purposes we 
build models with clumpy and
subvirial as well as equilibrium
initial conditions to investigate the correspondence of such models with observations of NGC~1818.  

All of our simulations that begin with stars initially distributed in clumpy density distributions
relax to smooth density distributions within the range of age estimates for NGC~1818 (see e.g., the bottom panel of Figure~\ref{xyplots}),
although some of our most highly substructured and subvirial simulations retain some minor clumpiness at an age of 15 Myr (which is erased by 30 Myr).

We list in Table~\ref{tab1} the initial virial ratio ($Q$), initial fractal dimension ($D$) and 
initial virial radius ($R_{\rm V}$) for the models that match the observed surface density distribution of NGC~1818 at the cluster's age, 
as well as the number of simulations ($N_{\rm sims}$) and the percentage of 
lines of sight to the simulations that match the observed radial distribution 
of the binary frequency ($P_{\rm obs}$, discussed in Sections~\ref{simtoobs}~and~\ref{discuss}). 
In the following, we will refer to a particular model using the convention of [$Q$,$D$].

\begin{deluxetable}{c c c c c}
\tablecaption{Summary Table of $N$-body Simulations\label{tab1}}
\tablehead{\colhead{$Q$} & \colhead{$D$} & \colhead{$R_{\rm V}$ (pc)} & \colhead{$N_{\rm sims}$} & \colhead{$P_{\rm obs}$ (\%)}}
\startdata
0.5 & 3.0 & 7  & 20 &  28.13 \\ 
0.5 & 2.6 & 8  & 2  &  99.3 \\ 
0.5 & 2.0 & 9  & 2  &  85.4 \\ 
0.5 & 1.6 & 11 & 2  &  78.0 \\
&&&&\\
0.3 & 3.0 & 7  & 10 &  17.03 \\
0.3 & 2.6 & 9  & 2  &  90.7 \\
0.3 & 2.0 & 10 & 2  &  53.3 \\
0.3 & 1.6 & 11 & 2  &  97.4 \\
&&&&\\
0.1 & 3.0 & 10 & 10 &  22.77 \\
0.1 & 2.6 & 10 & 2  &  94.7 \\
0.1 & 2.0 & 11 & 2  &  50.5 \\
0.1 & 1.6 & 11 & 2  &  20.3 \\
\enddata
\tablenotetext{}{
\footnotesize
$Q$ is the initial virial ratio.
$D$ is the initial fractal dimension (with $D$~=~3.0 corresponding to non-fractal, smooth, initial conditions).
$R_{\rm V}$ is the initial virial radius of the Plummer model.
$N_{\rm sims}$ is the number of simulations with the given [$Q$,$D$,$R_{\rm V}$].
$P_{\rm obs}$ is the percentage of sight lines that reproduce the observed radial distribution of the binary frequency.} \\
\end{deluxetable}

NGC~1818 is located at about 3.8 degrees ($\sim$3.3 kpc) from the center of the LMC.
We simulate the effects of the cluster orbiting within the potential of the LMC by placing the cluster at 3.3 kpc from the center of a point 
mass of $10^{10}$ \Msolar~on a linearized circular orbit.  Although the modeling of the cluster's orbit could be more sophisticated,
precise cluster orbital parameters are currently unknown and, more importantly, after only 30 Myr (the likely maximum age of NGC~1818) the effects on 
the cluster from the LMC tidal field are minimal.

Observations of NGC~1818 suggest a total binary frequency of up to 100\% \citep[at least for F-type stars;][]{hu10}.
We choose to begin all of our models with a 100\% binary frequency with no radial or primary mass dependence.  
We follow empirically defined initial distributions for the orbital parameters of the binaries in 
our models, as in \citet{gel13}, which agree with observations of solar-type binaries in young open clusters \citep[e.g. M35;][]{gel10}
and the field \citep{rag10}.  

Specifically, we draw the initial binary orbital periods from a
log normal distribution with a mean of $\log (P$~[days])~=~5.03 and $\sigma = 2.28$.  We allow the initial 
binaries to populate the entire log normal distribution, even though some of these binaries are born soft, because we are interested in 
investigating the dynamical disruption of soft binaries within the simulations.  In practice, the initial log normal 
distribution allows binaries with periods up to $\sim10^{10}$ days.
Consequently, a small fraction of binaries ($<$1\%) are assigned orbital separations that are larger than the distance
from the binary's center of mass to its nearest star.  Such binaries would likely not form in real clusters, and
are promptly disrupted dynamically at the start of the simulation, causing a drop in the binary frequency of about 0.2\% to 5\% 
inside of one $r_{\rm h}(0)$ and about 0.2\% to 2\% outside of one $r_{\rm h}(0)$ (depending on the initial density distribution).
This decrease is similar in magnitude to the uncertainty in the binary frequency 
from Poisson counting statistics (of about 1\% inside or outside of $r_{\rm h}(0)$), and is small compared to the $\sim$30\% difference in binary frequencies 
of the inner- and outer-most bins observed by \citet{deg13} and \citet{li13} in NGC~1818.

The NGC~1818 F-type binaries with mass ratios $q \geq 0.55$ have a mass-ratio distribution consistent with
${\rm d}N /{\rm d}q \propto q^{-\alpha}$, where $\alpha=0.4$ or $\alpha=0.0$ \citep{deg13,li13}.
In our simulations, we define the initial mass ratios by first taking two masses, $M_1$ and $M_2$,
from the \citet{kro01} IMF (within the mass limits defined above), combining these masses ($M_{\rm tot}$~=~$M_1$~+~$M_2$),
and then choosing a new mass ratio from a uniform distribution such that 
the new primary and secondary masses sum to equal 
$M_{\rm tot}$, and the new secondary mass 
is greater than 0.1~\Msolar\ but less than the new primary mass.
For the mass range observed by \cite{deg13} and \citet{li13}, this procedure produces a mass-ratio distribution (over all $q$) that favors low $q$ values
and approximately follows an $\alpha=0.4$ distribution (as also suggested by \citealt{kou05,kou07} and \citealt{reg11} for binaries of a range of spectral types).
Finally, we draw the initial eccentricities from a Gaussian distribution with a mean of $e = 0.38$ with $\sigma = 0.23$.
These binaries are then evolved through the \citet{kro95b} pre-main-sequence evolution prescription, with the same modifications 
as in \citet{gel13}.

We produce multiple realizations of each simulation using different initial random seed values to address the stochastic effects present in $N$-body simulations.
Moreover, for all simulations with a given [$Q$,$D$], the initial stellar and binary parameters (e.g., positions, velocities, 
binary periods and mass ratios, etc.) are all drawn from the same respective distributions, but 
each simulation randomizes the parameters to produce a unique initial stellar population.  
For the smooth (fractal dimension $D = 3.0$) models, we ran 20 $Q = 0.5$ simulations and 10 simulations each for 
$Q = 0.3$ and 0.1.  For the substructured models with a fractal dimension $D < 3.0$, we ran two simulations each.
We discuss our method for combining the results from these simulations below.
Our primary goal here is not to identify the specific combination of [$Q,D$] with which NGC~1818 most likely formed, but 
rather to investigate whether we can reproduce the observations of NGC~1818 with virial or subvirial and smooth or clumpy initial 
conditions.  As we show below, this number of simulations is sufficient to answer this question.

\section{Dynamical Evolution of the Binary Frequency} \label{phenom}

\begin{figure}[!t]
\epsscale{1.0}
\plotone{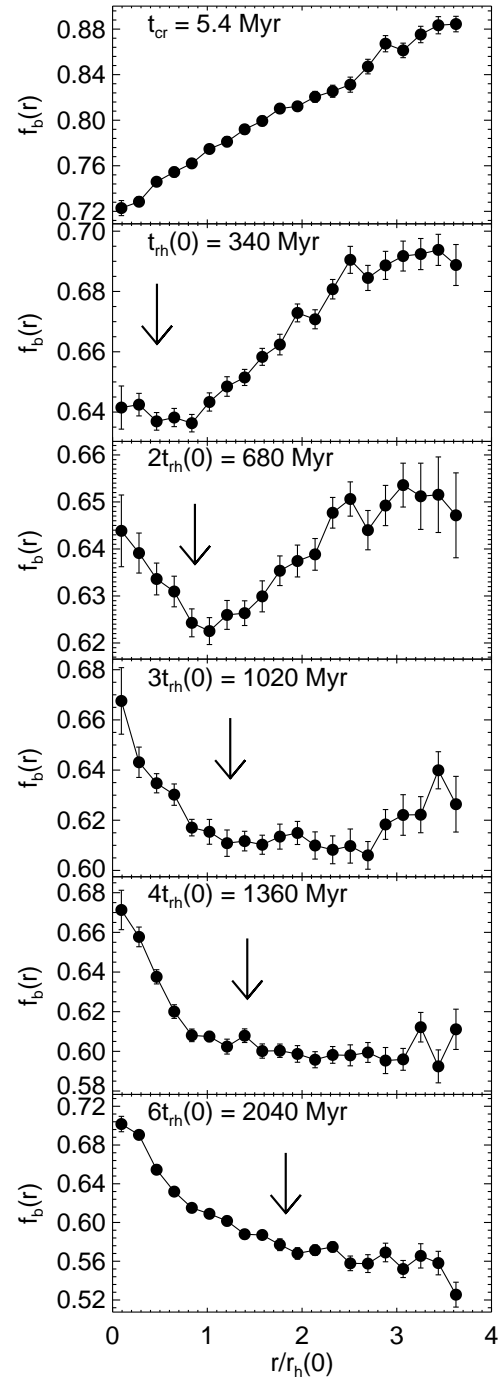}
\caption{
Binary frequency as a function of radius from the center of the [0.5,3.0] cluster model. Each point plots the mean binary frequency 
over all primary masses and mass ratios within a three-dimensional shell centered at the given radius and with a width of 1 pc ($\sim 0.19$~$r_{\rm h}(0)$) derived 
from the 20 realizations of our [0.5,3.0] cluster model.  The uncertainty at 
each point shows one standard error of the mean.  We show 
six snapshots of the cluster evolution here, labeled within 
each panel, namely, the crossing time ($t_{\rm cr}$~=~5.4~Myr, top) and multiples of the initial half-mass relaxation time ($t_{\rm rh}(0)$~=~340~Myr, bottom five panels).
The arrows in the bottom five panels mark the respective cluster radii
inside of which the local dynamical friction timescale 
for a binary with a mass equivalent to that of the mean binary mass in the cluster is shorter than the simulated time.
(At $t_{\rm cr}$ the arrow would be at a radius of $\sim$0, and is therefore not shown.)
\label{fbvr}
}
\end{figure}

Before we compare the $N$-body simulations to the observations of NGC~1818, we discuss in general the evolution of 
the radial distribution of the binary frequency.  
We choose here to focus on the 20 [0.5,3.0] (equilibrium and smooth) simulations, although the general evolutionary sequence we 
discuss below is common to all of our models.  
In Figure~\ref{fbvr} we show the binary frequency (over all primary masses and binary mass ratios) as a function of radius from the cluster center.  
We show six
representative times in the figure, namely the crossing time, $t_{\rm cr} \sim 5.4$~Myr, and multiples of the initial half-mass
relaxation time, $t_{\rm rh}(0) \sim 340$~Myr. (Although not plotted in this figure, the initial binary frequency has no radial dependence.)

\begin{figure}[!t]
\epsscale{1.0}
\plotone{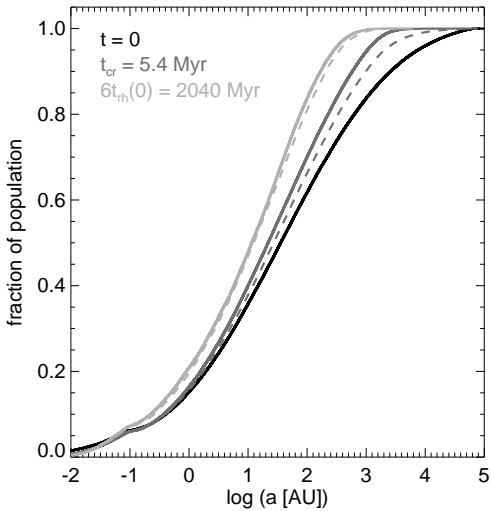}
\caption{
\label{afreq}
Cumulative semi-major axis distributions for binaries within the [0.5,3.0] simulations.  Each line shows the combined cumulative distribution 
including binaries from all 20 simulations.  We show three different snapshots of the cluster evolution: at $t$~=~0 (black, the initial distribution),
at the crossing time ($t_{\rm cr}$~=~5.4~Myr, dark gray), and at six times the initial half-mass relaxation time (6$t_{\rm rh}(0)$~=~2040~Myr, light gray).  For each time, the solid line 
plots the semi-major axis distribution for binaries within one initial half-mass radius ($r_{\rm h}(0)$~=~5.38~pc), and the dashed line plots the distribution 
for binaries outside of one $r_{\rm h}(0)$.  At $t$~=~0 the two lines overlap, as we do not impose any radial dependence on the initial conditions for binary orbital 
parameters. (The small ``bump'' in the distributions near $\log a$~=~-1.0 derives from the \citet{kro95b} pre-main-sequence evolution prescription.) 
}
\end{figure}

\begin{figure}[!t]
\epsscale{1.0}
\plotone{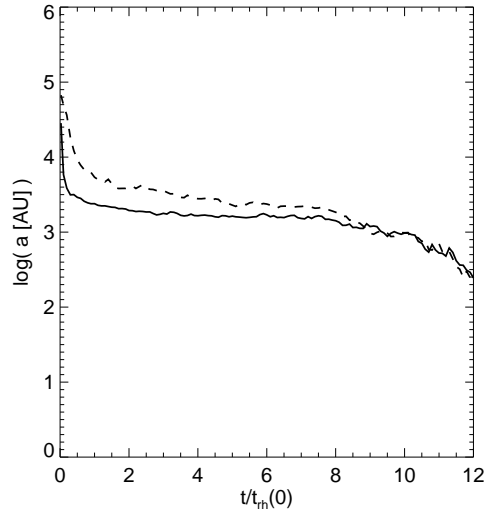}
\caption{
\label{mavtime}
Maximum semi-major axis of binaries inside (solid line) and outside (dashed line) of one initial half-mass radius ($r_{\rm h}(0)$~=~5.38~pc) within the [0.5,3.0] simulations as 
functions of time in units of the initial half-mass relaxation time ($t_{\rm rh}(0)$~=~340~Myr).  
At each time, we show the mean of the maximum semi-major axes within the given radial bin from the 20 simulations.
}
\end{figure}

After one crossing time, the binary frequency decreases towards the core of the cluster, due to the disruption of wide binaries.  
This is illustrated further in Figure~\ref{afreq}, where we show the cumulative distributions 
of semi-major axes for binaries inside and outside of one initial half-mass radius ($r_{\rm h}(0)$) at a crossing time.  Compared to the initial 
semi-major axis distribution, both distributions at $t_{\rm cr}$ are truncated at shorter separations by disruptive stellar encounters.
The very wide binaries observed in the field are soft even in the cluster halo.  
Importantly, the semi-major axis distribution of binaries found inside of one $r_{\rm h}(0)$ is shifted to even shorter separations than that of binaries
outside of one $r_{\rm h}(0)$.  This is also seen in Figure~\ref{mavtime}, where we show the maximum separation for binaries inside and 
outside  of $r_{\rm h}(0)$ as functions of time.  There is a steep drop in the maximum semi-major axis very early on due to the rapid disruption 
of very soft binaries from the primordial population in both the inner and outer cluster regions.  
However, the binaries outside of $r_{\rm h}(0)$ are able to retain companions at much wider separations than can survive inside of $r_{\rm h}(0)$, up to about 8~$t_{\rm rh}(0)$.

This difference between the inner and outer binaries results from the higher velocity dispersion and higher density in the cluster core relative to the halo.  
The higher velocity dispersion leads binaries in the core to move more rapidly, on average, relative to other stars than do binaries in the halo. 
Thus encounters within the core are more energetic and can disrupt tighter binaries.  
Also, the higher density results in a higher encounter rate in the core than in the halo \citep{lei11}.  
Furthermore, early in the cluster's evolution, the stars do not have sufficient time to mix throughout the cluster,
and instead most experience the dynamical environment near where they were born.  
These effects combine to produce a decreasing binary frequency towards the cluster core at $t_{\rm cr}$, which is maintained for roughly one $t_{\rm rh}(0)$.

As time progresses in the simulations, cluster-wide 
mass segregation effects begin to control the radial dependence of the binary frequency.  
The arrows in the five
lower panels in Figure~\ref{fbvr} mark the theoretical ``$r_{\rm min}$'' values or ``zones of avoidance'' 
from, e.g., \citet{map04} and \citet{fer12}.
In a given panel, $r_{\rm min}$ represents the radius inside of which the local dynamical friction timescale \citep{bin87} 
for a binary with a mass equivalent to that of the mean binary mass in the cluster is shorter than the simulated time.
Qualitatively, the $r_{\rm min}$ value predicts
the radius inside of which the binaries should experience the effects of dynamical friction and therefore fall towards the center of the cluster.
The result of this process is to increase the binary frequency in the core at the expense of the binary frequency towards the cluster halo.

At $2 t_{\rm rh}(0)$, the radial dependence of the binary frequency is high in the core, then drops to 
a minimum near $r_{\rm min}$ (at roughly one $r_{\rm h}(0)$), and then rises again towards the halo.  
The rate of dynamical binary (and triple-system) creation in our models is far too low to account for this increase in the binary fraction \citep[cf.][]{li13}.
Instead, this phenomenon is due to mass segregation processes.
Binaries inside of $r_{\rm min}$ have already fallen towards the core as the cluster 
begins to become mass segregated.  However binaries outside of $r_{\rm min}$ have not experienced enough dynamical friction to migrate 
fully towards the core.  
Also, at $2 t_{\rm rh}(0)$, the binary frequency at each radial bin has decreased to well below the values at $t_{\rm cr}$. By this time
more disruptive encounters have occurred locally, and also binaries that formed in the halo have had more time to 
orbit through the denser central regions of the cluster where encounters are more energetic.

Over time, the $r_{\rm min}$ value marches out towards the halo.  By
$4 t_{\rm rh}(0)$ the binary frequency maintains the rising distribution towards the core inside of $r_{\rm min}$ but develops a roughly flat distribution outside of $r_{\rm min}$.
By $6 t_{\rm rh}(0)$ the binary frequency increases continuously from the halo to the core of the cluster.
Also at $t \gtrsim 6 t_{\rm rh}(0)$, both the binaries inside and outside of $r_{\rm h}(0)$ have experienced sufficient encounters to regain similar
semi-major axis distributions (see Figures~\ref{afreq}~and~\ref{mavtime}).

As a check for the dependence of these results on the long-period cutoff of the initial binary orbital period distribution, we reanalyzed the [0.5,3.0] 
simulations considering all binaries with initial orbital periods $P>10^8$ days as single stars.
This reanalysis results in the same binary frequency in the core of the cluster at $t_{\rm cr}$ (and 30 Myr), and a progressively lower binary frequency towards 
the halo, as compared to the original analysis.
Binaries in the core with $P>10^8$ days are already broken up by one crossing time, where the binaries in the halo are not.  
The decreasing trend in binary frequency towards the core of the cluster remains.
After one $t_{\rm rh}(0)$ the radial distributions of the binary frequency from both analyses are indistinguishable, since nearly all of these wide binaries have been 
disrupted, even in the halo.

\section{Comparison Between Observations and Simulations of NGC~1818} \label{simtoobs}

Now we compare the simulations directly with observations of NGC~1818.  Here we choose to investigate individual simulations
rather than combining the different simulations for a given [$Q,D$] pair, as in the previous section.  
We will focus specifically on three
simulations defined by [0.5,3.0],  [0.3,2.0], and  
[0.1,2.6], and with initial conditions that lead to a
radial dependence of the binary frequency 
that matches particularly well the observations from \citet{deg13} and \citet{li13} at the age of NGC~1818.

We choose to compare the simulations at an age of 30~Myr 
to the observations.
At 15~Myr (the minimum age estimate for the cluster), the surface density profiles for these simulations are nearly identical to those at 30~Myr, 
although the subvirial simulations have a slightly higher central surface density, a residual of the early collapse and subsequent restructuring that 
occurs on roughly a crossing time.  The binary frequencies in these three simulations are also slightly higher at 15~Myr than at 30~Myr
(although still consistent with the observations), since the binaries have had less time to undergo disruptive dynamical encounters with other stars. 

At 30 Myr, the total masses for the [0.5,3.0], [0.3,2.0] and  [0.1,2.6]
simulations are 19,518.6 \Msolar, 19,064.4 \Msolar\ and 17,664.8 \Msolar, respectively.
For reference, the mean masses for all simulations with these respective [$Q,D$] pairings are 
19,550~$\pm$~140~\Msolar, 19,160~$\pm$~100~\Msolar\ and 18,046~$\pm$~380~\Msolar\ 
(showing uncertainties of one standard deviation for the $D=3.0$ simulations and half of the range in masses for the two simulations with $D<3$).
As the initial virial ratio decreases, the simulated clusters lose more mass early on, due to both violent relaxation processes that can eject more 
stars from the core of the cluster and 
the more loosely bound halo stars that are more vulnerable to tidal stripping (as the $Q$=0.1 and $Q$=0.3 simulations began with an initial virial radius of 10~pc, compared
to 7~pc for the $Q$=0.5 simulation).  These simulation masses agree well with the range of estimates from \citet{deg02} 
and \citet{mac03}.

At 30 Myr, the total binary frequencies in the [0.5,3.0], [0.3,2.0], and  [0.1,2.6]
simulations are 73\%, 78\%, and 80\%, respectively.
For reference, the mean binary frequencies for all simulations with these respective [$Q,D$] pairings are
73\%~$\pm$~1\%, 77.9\%~$\pm$~0.1\%, and 79\%~$\pm$~1\%
(with uncertainties defined as above for the masses). 
Again, these values agree well with the observational estimates for NGC~1818.

\begin{figure}[!t]
\epsscale{1.0}
\plotone{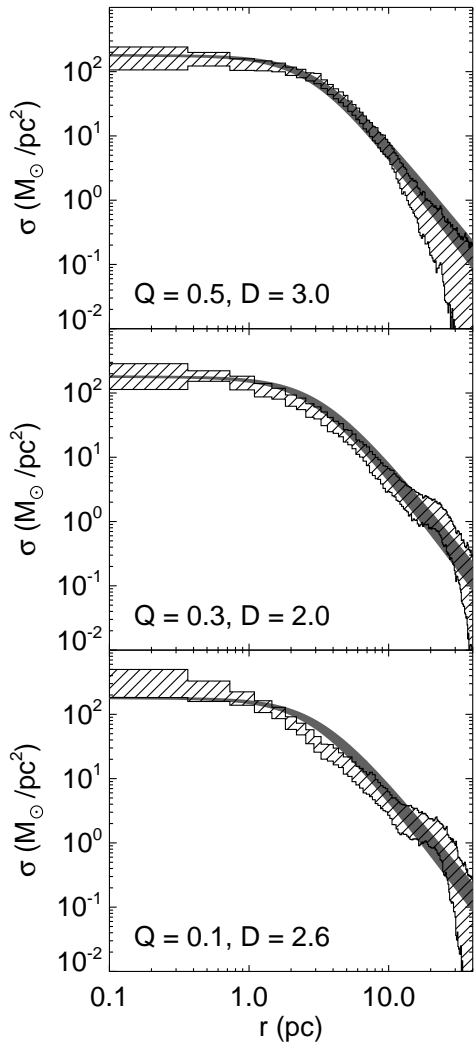}
\caption{Projected radial mass surface density profiles for three simulations compared with the EFF profile fit to observations of NGC~1818 by 
\citet{mac03}.  We show results from three specific simulations at 30 Myr defined by
[0.5,3.0] (top panel),
[0.3,2.0] (middle panel)
[0.1,2.6] (bottom panel) with the hatched regions.
Each bin for the respective simulations shows the range within which fall 95\% of our 1000 random sight lines. The solid gray
band shows the region encompassed by the \citet{mac03} EFF model, with parameters $\log \mu_0$~=~3.35~$\pm$~0.02~\Lsolar~pc$^{-2}$, $a$~=~52~$\pm$~3~pc,
and $\gamma$~=~2.76~$\pm$~0.12. 
\label{sdens}
}
\end{figure}

Next we investigate each simulation in projection, as an observer of NGC~1818 would only have access to two dimensions of spatial data.
Strictly, the orbit of the simulated cluster within the LMC potential defines a preferred line-of-sight projection for the cluster relative 
to an observer on Earth.  However, given the uncertainties in the true cluster orbital parameters and our simplistic modeling of the 
LMC potential, we choose not to define a particular projection to the cluster.  Furthermore, asymmetries within the simulations beginning with 
cool and clumpy conditions are not expected to exactly reproduce the initial cluster conditions, and are only meant as statistical 
approximations.  Therefore we select 1000 different lines of sight randomly distributed over the sphere and project each simulation along these 
sight lines.  In the following we combine these results to show the range in possible observed projections for a cluster like NGC~1818.

In Figure~\ref{sdens}, we show the 30 Myr mass surface density profile for these three simulations compared with the EFF model fit to the observations
by \citet{mac03}.  Although these simulations began with Plummer profiles, all simulations reproduce the observed surface 
density profile to within the range from our 1000 sight lines out to the $\sim$20 pc limit of the \citet{deg13} and \citet{li13} observations.  These results are not 
particularly sensitive to the chosen projection, nor are they sensitive to the specific simulation chosen from these sets.
Importantly, reproducing the observed surface density profile, total mass, and binary frequency indicates that we have correctly modeled the dynamical environment 
for binaries in the real cluster within our simulations.

\begin{figure}[!t]
\epsscale{0.95}
\plotone{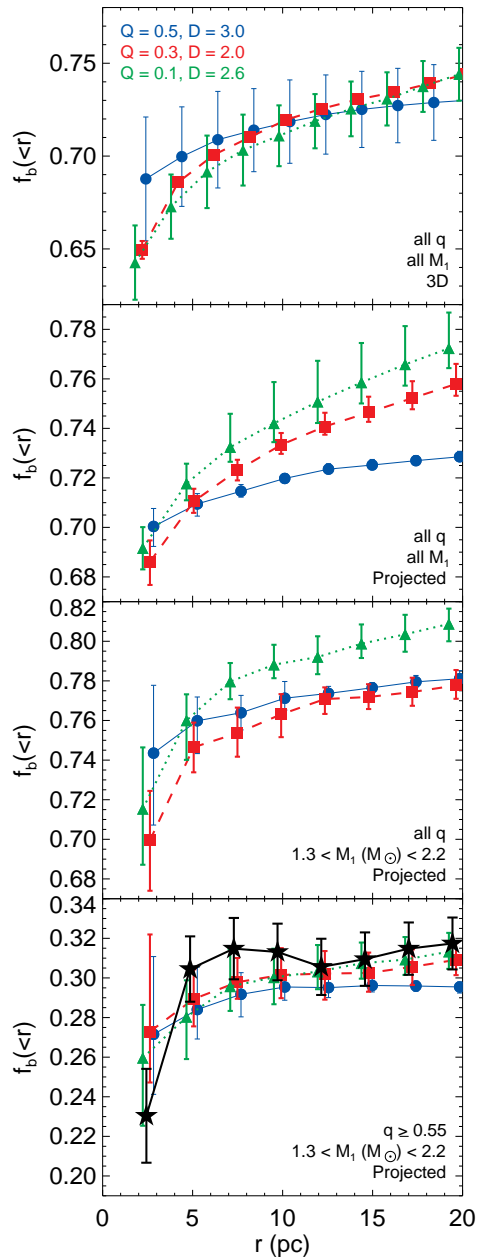}
\caption{
Cumulative binary frequency as a function of radius from the cluster center comparing results from simulations with
[0.5,3.0] (black/blue circles and solid lines),
[0.3,2.0] (dark-gray/red squares and dashed lines), and
[0.1,2.6] (light-gray/green triangles and dotted lines) 
to the \citet{li13} observations (black stars and solid line in the bottom panel).
In the top panel, each point plots the mean binary frequency inside of the given radius derived from all simulations 
of the respective [$Q,D$] pairing, and
the error bars show half of the range in binary frequency across all simulation of the given [$Q$,$D$] pairing in the given bin.
The bottom three panels show the three specific simulations 
in projection;  each point plots the mean 
binary frequency inside of the given radius out of the 1000 different line-of-sight projections, and the uncertainties show the range within which lie 95\% 
of the projections. 
The binary sample selection criteria in primary mass ($M_1$) and mass ratio ($q$) are written in each panel.
The observations assume ${\rm d}N /{\rm d}q \propto q^{-0.4}$, and error bars show $1\sigma$ uncertainties. 
All samples use the same radial bins, but we shift the simulation results slightly along the $x$ axis here for clarity.
(A color version of this figure is available in the online journal.)
\label{fbvrobs}
}
\end{figure}

In Figure~\ref{fbvrobs} we investigate the cumulative radial distribution of the binary frequency at 30 Myr in the three simulations and compare these 
results to the observations in the bottom panel.  Note that here we plot the results in cumulative form rather than binned as in Figure~\ref{fbvr}, to match 
the analysis of the observations.  Moreover, here each point plots the binary frequency inside the given radius.  For comparison,
we plot in the top panel mean results from all simulations in the respective [$Q,D$] pairs in three-dimensional shells
without limiting the simulations by primary mass or mass ratio.  
The [0.5,3.0] and [0.1,2.6] simulations show a relatively large
range in binary frequencies at each bin between the individual simulations with the given [$Q$,$D$] combination.
We will return to this below.

The remaining three panels show only the individual simulations we focus on here (and which are also shown in Figure~\ref{sdens}). In the second panel from the 
top, we show cumulative binary frequency distributions from these three simulations in projection for binaries of all primary masses and mass ratios.  
A visual comparison of the uncertainties plotted in this panel (which show the range within which lie 95\% of the projections) reveals that
the [0.5,3.0] simulation is most symmetric at 30 Myr, followed by the 
[0.3,2.0]
simulation, and finally the 
[0.1,2.6]
simulation.  

In the bottom two panels, we only include stars with primary masses between 1.3 \Msolar~and 2.2 \Msolar,
matching the mass range from the \citet{li13}
observations\footnote{Because the observations were selected to be within a range in magnitude, some near equal-mass binaries with
primary masses $<1.3$~\Msolar\ are included in the observational sample that are not included in the analysis of the simulations, 
and the opposite is true at the bright end of the observed magnitude range for primary masses $>2.2$~\Msolar.  
We assume that the contributions to the observed binary frequency from binaries gained and lost at the faint and bright ends of the magnitude range, 
respectively, approximately cancel each other out.}.  
Here, in general, the uncertainties 
become larger towards the cluster core
relative to the distributions in projection using all primary masses and mass ratios.
The increased uncertainties are primarily due to Poisson noise owing to the decreased number of objects when selecting only the observed regime 
in primary mass (and also mass ratio in the bottom panel).

Finally in the bottom panel we further 
limit the simulations to only include binaries with mass ratios $q \geq 0.55$ (and count binaries with $q < 0.55$ as single stars in our calculations of the 
binary frequencies).  In this panel we compare the simulations directly to the observations from \citet{li13}. All simulations closely reproduce the observations.

To test the relative difference between the observations and a given simulation, 
we performed $\chi^2$ tests comparing all 1000 lines of sight to the three simulations shown in Figure~\ref{fbvrobs} against the observations (bottom panel).
Specifically, for each line of sight, we calculate 
\begin{equation}
\chi^2 = \sum\limits_r \frac{(f_{\rm s}(<r)-f_{\rm o}(<r))^2}{e[f_{\rm o}(<r)]^2}, 
\end{equation}
where $f_{\rm s}(<r)$ and $f_{\rm o}(<r)$ are the simulated
and observed cumulative binary frequency inside the given radius, respectively, and $e[f_{\rm o}(<r)]$ is the error on the observed cumulative binary frequency 
at the same radius.  We have not constrained the simulation results, and therefore we have eight degrees of freedom (one for each bin in Figure~\ref{fbvrobs}).
We find that, respectively, 
97.4\%, 99.9\%, and 100\%
of the [0.5,3.0], [0.3,2.0], and [0.1,2.6]
projections are indistinguishable 
from the observations (at the $3 \sigma$ level).
Thus the specific line of sight for these simulations does not change the correspondence with the observations.

In Table~\ref{tab1} we show all [$Q,D$] pairs that reproduce the observed surface density profile of NGC~1818 at 30 Myr, and also the combined percentages of 
sight lines to these simulations that result in a trend in binary frequency with radius that is indistinguishable from the observations ($P_{\rm obs}$).
Specifically, we perform $\chi^2$ tests comparing the \citet{li13} observations to all 1000 sight lines towards a given simulation
as described above.
Then $P_{\rm obs}$ is defined as the percentage of sight lines to all simulations for a given [$Q,D$] pairing which result in a $p$-value of less than $3 \sigma$.

\section{Discussion} \label{discuss}

All of the simulations in Table~\ref{tab1} reproduce the observed surface density profile of NGC~1818 at 30 Myr, and
certainly there are more combinations of initial [$Q$,$D$,$R_{\rm V}$] that will also reproduce these observations.
About 60\% of the sight lines to these simulations (the mean of the 12 $P_{\rm obs}$ values in Table~\ref{tab1}) also reproduce the observed radial trend in binary frequency. 
Furthermore, if we take the average value at each radius of the cumulative binary frequency at all 1000 projections 
for all simulations of a given initial [$Q$,$D$,$R_{\rm V}$], we find that two thirds of these sets of initial conditions evolve to be indistinguishable from the observations at 30 Myr.
Moreover, the available observations do not constrain the $N$-body simulations to one particular set of initial conditions.  
Instead, it appears that a wide range of initial virial ratios and fractal dimensions can reproduce the observations at the age of NGC~1818.

Although our primary goal here is not to define the precise primordial conditions with which NGC~1818 was likely born, the $P_{\rm obs}$ values we calculate 
here for our simulations do indicate a preference towards 
clumpy initial conditions.  The mean $P_{\rm obs}$ values for the 
four respective $Q$ = 0.1, 0.3, and 0.5 models at all $D$ values are 
47\%~$\pm$~17\%, 65\%~$\pm$~19\%, and 73\%~$\pm$~16\%.
Given the relatively large range in $P_{\rm obs}$ values for the simulations at each value of $Q$, 
we do not detect a preferred virial ratio for reproducing the observations.
However, the mean $P_{\rm obs}$ values for the 
three respective models with $D$ = 3.0, 2.6, 2.0, and 1.6 at all $Q$ values are 23\%~$\pm$~3\%, 95\%~$\pm$~3\%, 63\%~$\pm$~11\%, and 65\%~$\pm$~23\%.
A clumpy primordial population 
reproduces the observations more closely than a
smooth primordial population.

As noted above, in some cases we see significant variations between simulations of the same [$Q,D$] pairing
(for example, see the uncertainties in top panel of Figure~\ref{fbvrobs}).
Furthermore, although two simulations of the same [$Q,D$] pairing may both reproduce the observed surface 
density profile, only one may reproduce the observed radial dependence of the binary frequency, and in many cases this difference cannot easily 
be attributed to projection effects (at least given our 1000 sight lines).  
Note that we include stars of all masses in the surface density profile, but only stars within the primary-mass and mass-ratio 
range of \citet{li13} in the radial dependence of the binary frequency,
which comprise $\sim$3\% of the total stellar population of the cluster at 30 Myr.
This may indicate that small asymmetries in the primordial population 
and Poisson noise are important for determining the 
radial dependence of the binary frequency at a few $t_{\rm cr}$, particularly when examining a 
magnitude- or mass-limited sample.  
The larger number of simulations beginning with smooth ($D=3$) initial density distributions alleviates much of this uncertainty, but
additional simulations may be required to determine very robust $P_{\rm obs}$ values for 
[$Q,D<3$] pairings.

In summary,
we confirm here the hypothesis of \citet{deg13} and \citet{li13} that the observed decreasing radial trend in binary frequency towards the core of NGC~1818 
can be reproduced by the early dynamical disruption of binaries by stellar encounters within the cluster.  This trend can be produced within clusters born both 
in equilibrium or collapsing as well as with smooth or clumpy primordial density distributions, although our analysis indicates a modest preference for 
clumpy primordial conditions for NGC~1818.

For a cluster that is born with a binary population that has similar distributions of orbital parameters to those of the solar-type binaries in the 
Galactic field (with no radial dependence),
the disruption of soft binaries naturally produces a decreasing trend in binary frequency towards the cluster core in roughly a crossing time.
As described in Section~\ref{phenom} and illustrated in Figures~\ref{fbvr}~to~\ref{mavtime}, 
the higher density and velocity dispersion in the core result in more frequent and
more energetic encounters that ionize wide binaries, while binaries of similar binding energies can survive in the less dense halo \citep[see also][]{sol08}. 
The formation of this type of early distribution relies on primordial soft binaries, and likely will not arise if, for example, the primordial binaries are 
cut off at the initial hard-soft boundary of the core.

At the age of NGC~1818, we find that our simulations that start with more clumpy and subvirial initial conditions (lower values of both $D$ and $Q$) 
have the largest differences between the binary frequencies inside and outside of $r_{\rm h}(0)$.
For example, the percent difference between the mean binary frequencies 
inside and outside of $r_{\rm h}(0)$ for the [0.1,1.6] simulations is 17\% at 30 Myr, while for the [0.5,3.0] simulations we find only a 7\% difference 
at the same age.  
This can also be seen in the top two panels of Figure~\ref{fbvrobs}, where the distribution for the [0.5,3.0] model is flatter than for those of 
the initially subvirial and clumpy simulations.  Clusters with more highly substructured and subvirial initial conditions undergo a more rapid 
relaxation process \citep[e.g.][]{mcm07} and, in turn, disrupt (wide) binaries more efficiently.

After roughly six initial half-mass relaxation times (about 2040~Myr)
both the core and halo binary populations have experienced sufficient encounters to
regain similar semi-major axis distributions.
Over this time, mass segregation processes become most important in determining the radial dependence of the binary frequency in the cluster.
Figure~\ref{fbvr} illustrates this turn-over from early dynamical disruption to mass segregation.  

Notably, the radial dependence of the binary frequency develops a minimum value after about 1~$t_{\rm rh}(0)$ ($\sim$340~Myr), which marches outwards towards the 
halo until about 4~$t_{\rm rh}(0)$ ($\sim$1360~Myr), when the outer cluster no longer maintains a second peak in binary frequency.
In our simulations this bimodal radial distribution in the binary frequency manifests because of 
both early binary disruption and subsequent
mass segregation processes. 
Furthermore, because the radial distribution of the binaries changes predictably with 
time (moving from decreasing towards the core, to a bimodal distribution, to rising towards the core), observations of the radial dependence
of the binary frequency in a cluster can help to constrain the cluster's dynamical age (e.g., the number of $t_{\rm rh}(0)$ that the cluster has lived through).

Interestingly, the similarly aged LMC star cluster NGC~1805 shows evidence for a minimum in the radial distribution of the binary frequency near 
a radius of about 15--20 arcsec \citep[about 3.6--4.9 pc;][]{li13}.
This may indicate that NGC 1805 is more \textit{dynamically} evolved than NGC~1818 despite having the same chronological age.
Indeed, the total masses of NGC 1805 and NGC~1818 of $\log(M_{\rm cl}$[\Msolar])~=~3.45 and 4.01 and the numbers of stars of 
6000 and 9000 \citep{li13}, and the core radii of 1.33~pc and 2.45~pc  \citep{mac03}, 
imply present-day half-mass relaxation times for NGC 1805 and NGC~1818 of about 115~Myr and 215~Myr, respectively.  Detailed models of NGC 1805 will be important
to understand if such a bimodal distribution can develop from an initial binary population born with no radial dependence in frequency (or distributions
of orbital parameters and masses) after only about one fourth of a half-mass relaxation time.

Bimodal radial distributions have also been observed previously for the frequency of blue stragglers in many globular clusters.
\citet{map04} and \citet{fer12} argue that such distributions are the result of dynamical friction and mass segregation processes acting 
on the binary progenitors of the blue stragglers, much like we see here for the binaries in our NGC~1818 simulations.
\citet{bec13} find that the blue stragglers and possibly also the binaries in the globular cluster NGC 5466 show 
bimodal radial distributions with minimum values located at approximately the same radial distance from the cluster center, a significant step towards confirming this hypothesis.
As noted by \citet{fer12}, the blue stragglers can also serve as an indicator of a cluster's dynamical age, assuming that these same mass segregation 
processes govern the evolution of the blue straggler radial distribution in a cluster.

Alternatively these observed bimodal blue straggler distributions may be the result of different formation processes, owing to the different densities in 
the different cluster regions \citep{fer97,li13a}. Dynamical ejections of blue stragglers from the cluster core, perhaps during 
their dynamical formation, may also contribute to forming such bimodal blue straggler radial distributions \citep{sig94}.

Finally, we note that \citet{els98} found the opposite radial trend for the binary frequency of NGC~1818 in their observations of binaries with primary masses 
between 2~\Msolar\ and 5.5~\Msolar\ and $q \gtrsim 0.7$.
Specifically they found a binary frequency of 35\%~$\pm$~5\% inside the core (using a core radius of $\sim$2 pc) which decreases to 
20\%~$\pm$~5\% outside of about 3 core radii.  \citet{deg13}
attribute this discrepant result to blending and the near-vertical morphology of the stellar main sequence within the magnitude range observed by \citet{els98}, which 
may have compromised their analysis of the observations. 
When we examine the observed mass range of \citet{els98} in our [0.5,3.0] simulations, all have significantly lower binary frequencies (for $q \geq 0.7$) in the core 
than the \citet{els98} value of 35\%~$\pm$~5\% (and note that we began our simulations with $f_{\rm b} = 100$\%), which adds further support to the conclusions of \citet{deg13} 
regarding the analysis of the observations.  

\citet{els98} also ran one $N$-body simulation to model the cluster binary population.  They were unable to simulate the full cluster or the high binary 
frequency due to computational limitations.  Instead they simulated a cluster with roughly half the expected initial mass of NGC~1818,
an initial binary frequency of about 17\%, and soft binaries were not included (4000 AU was the upper limit to the binary separations). 
They drew stellar positions and velocities from a smooth Plummer model in virial equilibrium.  
In their simulation, they found marginal evidence for an increase in the binary frequency within the core as compared to the halo, but they note that
given the relatively small sample sizes the error bars on the binary frequency at different radial bins were sufficient to allow for a flat distribution as well.
These early models showed that mass segregation could be effective in the mass range of the \citet{els98} sample, potentially explaining a rise in the 
binary frequency towards the core. However, given the indirect nature of the models and the exclusion of the soft end of the binary distribution, the
ability to reproduce the \citet{els98} observations with $N$-body models should be revisited, and particularly to understand if both the \citet{els98} and \citet{li13} results 
can be accommodated within one cohesive direct $N$-body simulation of NGC~1818 (which we will address in a future paper).

\vspace{1em}
\section{Conclusions} \label{conc}

NGC~1818 is a very valuable star cluster for our understanding of the early evolution of the binaries in a dense stellar environment.  The cluster 
exhibits a rare radial dependence of the binary frequency that decreases towards the core of the cluster, whereas many other (generally older) 
star clusters show the opposite trend \citep[e.g.][]{mat86,gel12,mil12}.  
Here we show, through a grid of sophisticated $N$-body simulations, that the observed surface density profile of NGC~1818 \citep{mac03} and 
the radial dependence of the binary frequency \citep{deg13,li13} can be reproduced simultaneously at the cluster's age by $N$-body simulations with both 
initially smooth or substructured and equilibrium or collapsing stellar populations, 
with a modest preference for substructured initial conditions.
The radial distribution of the binary frequency in a rich star cluster can transition smoothly 
over time from a uniform primordial radial distribution, to one that decreases towards the core at early ages, to one that rises towards the core at later ages.
Thus both rising and falling radial distributions in binary frequency can arise naturally from the evolution of a binary population within the 
same rich star cluster as a consequence of both dynamical disruption and mass segregation of the binaries.

\acknowledgments
Thanks to Andreas K\"upper for explaining the methods used in his McLuster code.  
This material is based upon work supported by a National Science Foundation Astronomy and Astrophysics Postdoctoral Fellowship under Award No.\ AST-1302765 and 
the Lindheimer Fellowship at Northwestern University.  
RdG and CL acknowledge partial research funding from the National Natural Science Foundation of China through grant 11073001.
RdG also acknowledges research support from the Royal Netherlands Academy of Arts and Sciences (KNAW) under its Visiting Professors Programme.

\bibliographystyle{apj}
\bibliography{ngc1818.ms}

\end{document}